\documentclass[12pt,a4paper]{article}

\def\w{v}
\usepackage{amsmath}
\usepackage{amsfonts} 
\usepackage{amssymb}
\usepackage[dvips]{graphicx}
\usepackage{epsfig}
\newcommand{\beq}{\begin{equation}}
\newcommand{\eeq}{\end{equation}}
\newcommand{\p}{\partial}
\newcommand{\bmm}{\begin{matrix}}
\newcommand{\emm}{\end{matrix}}
\newcommand{\bpm}{\begin{pmatrix}}
\newcommand{\epm}{\end{pmatrix}}
\newcommand{\bc}{\begin{center}}
\newcommand{\ec}{\end{center}}
\begin{document}
\begin{center}
{\Large \bf Microscopic description of 2d topological phases, duality and 3d state sums}\\
\vspace{30pt}

{\sl Zolt\'an K\'ad\'ar$^{1}$,  Annalisa Marzuoli$^{1,2}$ {\rm and} Mario Rasetti$^{1,3}$}\\

\vspace{24pt}

{\footnotesize
%\address{
$^1$Institute for Scientific Interchange Foundation,\\
Villa Gualino, Viale Settimio Severo 75, 10131 Torino (Italy)\\ 
{\tt email:} {\tt kadar@isi.it}\\
\vspace{3mm}
$^2$Dipartimento di Fisica Nucleare e Teorica, Universita degli Studi di Pavia, \\
Istituto Nazionale di Fisica Nucleare, Sezione di Pavia, via A. Bassi 6, 27100 Pavia (Italy)\\
{\tt email:} {\tt Annalisa.Marzuoli@pv.infn.it}\\
\vspace{3mm}
$^3$Dipartimento di Fisica, Politecnico di Torino, corso Duca degli Abruzzi 24, 10129 Torino
(Italy)\\ {\tt email:} {\tt mario.rasetti@polito.it}
%}
}
\end{center}
\begin{abstract}
Doubled topological phases introduced by Kitaev, Levin and Wen supported on two dimensional lattices 
are Hamiltonian versions of three dimensional topological quantum field theories described by the Turaev-Viro state sum
models. We introduce the latter with an emphasis on obtaining them from theories in the continuum. Equivalence of
the previous models in the ground state are shown in case of the honeycomb lattice and the gauge group being a 
finite group by means of the well-known duality transformation between the group algebra and the spin network 
basis of lattice gauge theory. An analysis of the ribbon operators describing excitations in both types of models 
and the three dimensional geometrical interpretation are given.    
\end{abstract}
%%%%%%%%%%%%%%%%%%%%%%%%%%%%%%%%%%%%%%%%%%%%%%%%%%%%%%%%%%%%%%%%%%%%%%%%%%%%%%%%%
\section{Introduction}
%%%%%%%%%%%%%%%%%%%%%%%%%%%%%%%%%%%%%%%%%%%%%%%%%%%%%%%%%%%%%%%%%%%%%%%%%%%%%%%%%
Topological quantum fields theories (TQFT) in three dimensions
describe a variety of physical and toy models in
many areas of modern physics. The absence of local degrees of freedom is a great simplification,
it often leads to complete solvability \cite{wittclass,Ati}. Perhaps the most recent territory, where they appeared to
describe real physical systems, is that of topological phases of matter, being e.g. responsible for 
the fractional quantum Hall effect \cite{wenold}. Since the idea of fault--tolerant quantum computation appeared in
the literature \cite{faulttor}, TQFT's are also important in quantum information theory. These new applications also 
enhanced the mathematical research, led to classification of the simplest models \cite{mtcc}.

Due to their topological nature, TQFT's admit discretization yet remaining an exact description of the theory
given by an action functional on a continuous manifold. One large class thereof are the 
so-called BF theories, whose Lagrangian density is given by the wedge product of a $(d-2)$ form  $B$ and the 
curvature $2$-form $F$ of a gauge field \cite{KaGoRa}. We will deal here with a special class of three dimensional theories, which
describe doubled topological phases and restrict our attention to discrete gauge groups $G$. 
The context they appeared in, in recent physics literature \cite{faulttor,levinwen}, are Hilbert spaces of states in two 
dimensions and dynamics therein, which are boundary Hilbert spaces $H$ of the relevant TQFT's. 
Operators acting on $H$ correspond to three dimensional amplitudes on the thickened surface. In this paper we will 
explain this correspondence, which was proved for the ground state projection recently \cite{prev}, and provide the geometric
interpretation of the ribbon operators, which create quasi--particle excitations from the ground state. This
is a step towards extending the correspondence to identify the ribbon operators as invariants of manifolds with 
coloured links embedded in them in the TQFT.  

The emergence of topological phases from a description of microscopic degrees of freedom are modelled by the lattice
models of Kitaev \cite{faulttor} and Levin and Wen \cite{levinwen}. Since they generically have degenerate ground states and 
quasi--particle excitations insensitive to local disturbances, they are also investigated in the theory of
quantum computation \cite{recrev}, their continuum limit being closely related to the spin network 
simulator \cite{compspinnet,freedman}. The ground states were 
extensively studied in the literature, their MERA (multi--scale entanglement renormalisation ansatz)
\cite{mera1,mera2} and tensor network 
representations \cite{tnr} have been constructed to study e.g. their entanglement properties \cite{raduetal,lw2}. Finding 
the explicit root of these structures in lattice gauge theory and TQFT can help to understand their physical properties.     

Lattice gauge theories admit seemingly very different descriptions. A state can be represented by
assigning elements of the gauge group to edges of the lattice. The dual description
in terms of spin network states, where edges are labelled by irreducible representations (irreps) of the gauge group 
and vertices by invariant intertwiners are also
well known since \cite{kogsuss}. To name an application, this description turned out to provide a convenient basis 
for most approaches to
modern quantum gravity theories \cite{rovellismolin,rovellibook}. In this paper we will show in detail how these 
dual descriptions give rise to Kitaev's quantum double models in one hand and the spin net models of Levin and Wen 
on the other. Then the ribbon operators in both models and their identification will be discussed. 

The organisation of the paper is as follows. In the next section, we introduce the Turaev--Viro models via the example
of BF theories. In section 3, we briefly introduce the string net models of Levin and Wen on the honeycomb lattice 
$\Gamma$ in the surface $S$ and recall the proof \cite{prev} that the ground state projection is given by the Turaev--Viro 
amplitude on $S\times [0,1]$. The boundary triangulations of $S$ are given by the dual graphs of $\Gamma$ 
decorated by the labels inherited from the "initial" and "final" spin nets. 
In section 4 the duality between the states of the Kitaev model and the string nets will be shown by  
changing the basis from the group algebra to the Fourier one. By using this duality and an additional projection, we will
obtain the electric constraint operators of the string net models. The matrix elements of the 
magnetic constraints are also recovered provided that the local rules of Levin and Wen hold. We explain that they do in all 
BF theories, which is a strong motivation in their favour  for the case when the gauge group is finite. 
In section 5, we discuss ribbon operators and give their three dimensional geometric interpretation in terms of 
framed links in the Turaev--Viro picture. Finally, a summary is given with a list of questions for future research.

%%%%%%%%%%%%%%%%%%%%%%%%%%%%%%%%%%%%%%%%%%%%%%%%%%%%%%%%%%%%%%%%%%%%%%%%%%%%%%%%
\section{Turaev--Viro models}
%%%%%%%%%%%%%%%%%%%%%%%%%%%%%%%%%%%%%%%%%%%%%%%%%%%%%%%%%%%%%%%%%%%%%%%%%%%%%%%%%
In three dimensions
both the $F(A)$ (the field strength) and $B$ fields of BF theory 
can be be considered to be forms valued in the Lie algebra of the gauge group $G$. The action can then be written as
$\int_M tr (B\wedge F)$ with $tr$ being an invariant non-degenerate bilinear form on the Lie algebra and $M$ is a smooth, oriented, 
closed three-manifold.
We may start  from the case when $G$ is a semi--simple Lie group relevant in
particle physics theories and gravity, $A$ being the connection
in the principal $G$--bundle over $M$. In three dimensions the "space--time" separated form of the Lagrangian has 
the structure $B_jdA_k+A_0D_jA_k+B_0F_{jk}(A)$ where $j,k$ are spatial indices\footnote{There is not necessarily 
physical time in the theory, one can do this decomposition for Euclidean signature as well.}. The first term is the
standard kinetic term, the second implies the (Gauss) constraint of gauge invariance, the third the vanishing 
of the (two-dimensional) field strength ($A_0$ and $B_0$ are Lagrange multipliers), $D$ and $d$ 
stand for the covariant and the exterior derivatives, respectively.  

Since locally the solution of the constraints is given by a pure gauge $A_i=\tilde{g}^{-1}\p_i \tilde{g}$ 
($\tilde{g}:S\to G$ smooth function, with $S$ being the spatial hypersurface) one may discretize the 
theory by introducing a lattice on the spatial surface and quantize the remaining degrees of freedom: the
holonomies (elements of $G$) describing the coordinate change between faces of the lattice. They correspond to the edges of the dual
lattice, which is constructed by placing a vertex inside each face and connecting new vertices, which were put inside 
neighbouring faces. This dual lattice is the starting point of the models in \cite{faulttor}, the electric
constraints are the remainders of the Gauss constraint and the magnetic ones are the remainders of the flatness 
constraint. For a detailed exposition see e.g. \cite{sympred}. 

%%%%%%%%%%%%%%%%%%%%%%%%%%%%%%%%%%%%%%%%%%%%%%%%%%%%%%%%%%%%%%%%%%%%%%%%%%%%%%%%%
%\subsection{String nets and the Turaev-Viro model} 
%%%%%%%%%%%%%%%%%%%%%%%%%%%%%%%%%%%%%%%%%%%%%%%%%%%%%%%%%%%%%%%%%%%%%%%%%%%%%%%%
The partition function of the above BF theory is formally obtained by taking the functional integral over 
the fields $A$ an $B$ of the phase associated
to the classical action 
\beq \label{ZBF}
{\bf Z}_{BF}=\int {\cal D}A\,{\cal D}B e^{i \int_M tr B\wedge F(A)}\ .\eeq 
It is not so easy to give this definition a precise sense, but for the moment it is not necessary
to go into further details. What does matter is that there exists
a consistent way to discretize the partition function by considering an oriented triangulation $M_\Delta$ of the manifold $M$
by assigning  two Lie algebra element $B_i,\Omega_i$ to each edge $i$ in $M_\Delta$. 
The generator $B_i$ can be thought as the integral of $B$ along
the edge $i$, whereas $\Omega_i$ as the logarithm of the group element corresponding to the holonomy around the edge
$i$\footnote{To be more precise, one needs to introduce the dual complex by putting vertices inside every tetrahedron, 
connecting those vertices which were put in neighbouring tetrahedra and a vertex should be singled out on the
boundary of each dual face. Then the procedure to get $\Omega_i$ is the following: Take the dual face corresponding 
to $i$. Multiply the holonomies along the boundary edges of this dual face starting from the vertex singled out  in 
a circular direction determined by the orientation of $i$ (say, by the right hand rule). The logarithm of this group element is $\Omega_i$.}. 
Then the Feynman integral in (\ref{ZBF}) can be replaced by 
$\prod_i\int dg_i\int dB_i e^{i\int tr\,B_i\Omega_i}$. The $B$ integrals will yield Dirac deltas 
$\delta_{g_i,{\bf 1}}$ and one can now proceed with decompositions  in terms of 
irreps of the gauge group. This way one ends up with a discrete state sum instead of the original Feynman integral, 
where each state is the
triangulation coloured with irreps and its weight is given by the precise final form of the amplitude (examples are given below). 
The structure of the partition function (amplitude) for a prototypical theory, the Ponzano--Regge model 
\cite{PoRe} corresponding to $G=SU(2)$, reads
\beq Z(M_\Delta)=\sum_{j_i}\prod_i d_{j_i} \prod_t (6j)_t\ ,\label{pr}\eeq
where $(6j)$ is the Wigner $6j$ symbol of $SU(2)$ depending on the $6$ irreps decorating the edges of the 
tetrahedron $t$, $d_{j_i}$ is the dimension of the irreps $j_i$ assigned to the edge $i$ and the sum ranges over 
all states, that is, all possible colourings of the edges with irreps. 
It turns out that this state sum is well defined and independent of the chosen triangulation for a large
class of models\footnote{This is one way to define a TQFT rigorously.}. For a systematic derivation of 
this state sum from action functionals, see \cite{kfa} or section 2.3 of \cite{danielet}.

The Ponzano--Regge partition function (\ref{pr}) is formally independent of $\Delta$, but is divergent. However, the
 Turaev-Viro (TV) model \cite{tv}, a regularized version thereof, has a well defined partition function, given by
\beq \mathbf{Z}_{\,TV}\,[M^3;q]\,=\,\sum_j\;d^{-V}\,
\prod_{i} d_{j_i}
\,\prod_t \;
\begin{Bmatrix}
j_1 & j_2 & j_3 \\
j_4 & j_5 & j_6
\end{Bmatrix}_t \label{tvpartf}\ ,\eeq
where the underlying algebraic structure is the quasi-triangular Hopf algebra $SU_q(2)$ with 
$q=\exp(2\pi i/k),\,k\in {\mathbb Z}$ fixed, $j\in [0,1,\dots,k-1]$ denotes the irreps of $SU_q(2)$, $d_j\in {\mathbb C}$ 
is the so-called quantum dimension of $j$, the constant $d$ is defined by $d=\sum_k d_k^2$, the quantity in the
brackets is the quantum $6j$ symbol and $V$ is the number of vertices of the triangulation. One finds the precise 
definitions of all the quantities along with the algebraic properties assuring consistency and triangulation independence \cite{PoRe} of
the amplitudes in \cite{tv}. We will briefly mention the origin of the latter property in the next section. 
The final fact for this introductory section is about the form of the amplitude (\ref{tvpartf}) for manifolds 
with non--empty boundary. The associated boundary triangulation, whose edges are
decorated by labels ${\{j'\}}$, derives from a given triangulation
in the  $3d$ bulk and is kept fixed. The amplitude reads
\beq \mathbf{Z}_{\,TV}\,[M^3, \{j'\};q]\,=\,\sum_{j\;int}\;d^{-V}\,
\prod_{i} d_{j_i}\prod_{i'}d_{j_{i'}}^{-\frac{1}{2}}\prod_{v'}d_{j_{v'}}
\,\prod_t \;
\begin{Bmatrix}
j_1 & j_2 & j_3 \\
j_4 & j_5 & j_6
\end{Bmatrix}_t \label{tvpartfb}\ ,\eeq
where $V$ is the number of internal vertices, the index $i$ ranges over internal edges, $i'$ over boundary edges, $v'$ 
over boundary vertices (each boundary vertex is the endpoint of an internal edge, $j_{v'}$ is its colour), $t$ 
over all tetrahedra and the summation is done for internal edge labels only, while those on the boundary are kept fixed.
 
Note that there is a quasi-triangular Hopf algebra associated to finite groups as well, the so-called Drinfeld 
(quantum) double ${\cal D}(G)$ \cite{Dri}. 
There, the dimensions $d_j$ as well as the $6j$ symbols can be obtained from the representation theory of the group $G$.  

%%%%%%%%%%%%%%%%%%%%%%%%%%%%%%%%%%%%%%%%%%%%%%%%%%%%%%%%%%%%%%%%%%%%%%%%%%%%%%%
%%%%%%%%%%%%%%%%%%%%%%%%%%%%%%%%%%%%%%%%%%%%%%%%%%%%%%%%%%%%%%%%%%%%%%%%%%%%%%%%
%%%%%%%%%%%%%%%%%%%%%%%%%%%%%%%%%%%%%%%%%%%%%%%%%%%%%%%%%%%%%%%%%%%%%%%%%%%%%%
\section{String nets}
%%%%%%%%%%%%%%%%%%%%%%%%%%%%%%%%%%%%%%%%%%%%%%%%%%%%%%%%%%%%%%%%%%%%%%%%%%%%%%

Levin and Wen \cite{levinwen} started off from the algebraic structure underlying the above models (consistent set of $6j$ symbols and
quantum dimensions), which serves as the algebraic data in defining TQFT's. Taking these data for granted, they constructed a two dimensional lattice
model, which we will now introduce briefly. Consider a surface $S$ with a fixed oriented honeycomb lattice $\Gamma$ embedded in it. 
The Hilbert space is spanned by all possible decorations
of the edges with labels $j$; we will refer to them as irreps (of $SU_q(2)$ or the finite group $G$) as we will not need 
to treat the most general TQFT's. The Hamiltonian is a sum of two families of mutually commuting constraint operators, 
\beq H=-\sum_v A_v-\sum_p B_p\ ,\label{ham} \eeq
where the first sum is over all vertices, the second is over all 
plaquettes of the lattice. $A_v\equiv N_{ijk}$ for $i,j,k$ being the
irreps decorating the edges adjacent to the vertex $v$ (the numbers $N_{ijk}\in {\mathbb N}$ are referred to as fusion
coefficients between the irreps: $i\otimes j=\sum_k N_{ijk}\, k$). The magnetic constraints are written as a sum 
\beq B_p=\frac{1}{d}\sum_s d_s\,B^s_p\label{mco}\eeq
over irreps and the action of the individual terms is
\beq\label{mag}\begin{array}{l}
B^s_{p} \,\;\Big|\;\bmm\includegraphics[height=0.8in]{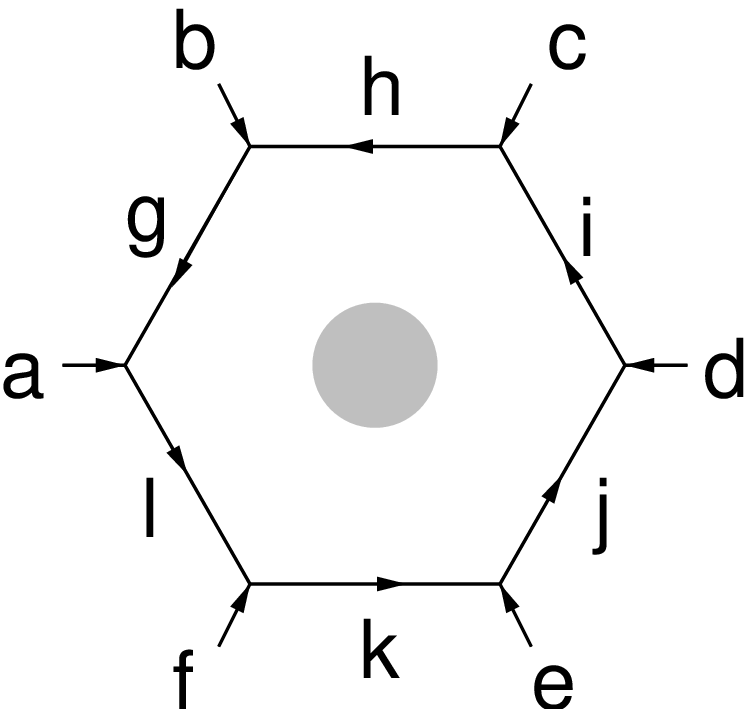}\emm\,\Big\rangle=\\
{\displaystyle\sum_{g'h'i'j'k'l'}
F^{bg^*h}_{s^*h'g'^*}
F^{ch^*i}_{s^*i'h'^*}
F^{di^*j}_{s^*j'i'^*}
F^{ej^*k}_{s^*k'j'^*}
F^{fk^*l}_{s^*l'k'^*}
F^{al^*g}_{s^*g'l'^*}
\;\Big|\;}
\bmm\includegraphics[height=0.8in]{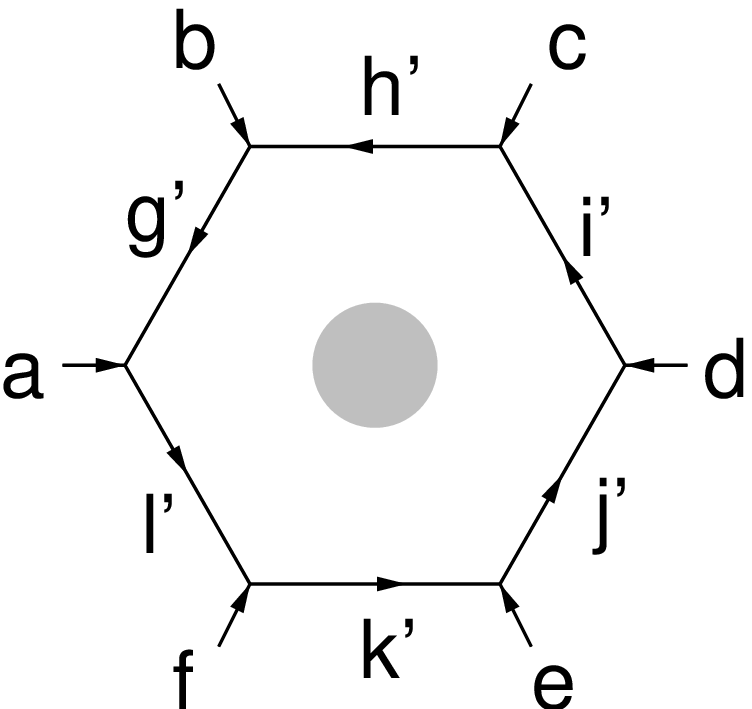}\emm\,\Big\rangle\ ,
\end{array}\eeq
while its action on the rest of the state supported on the honeycomb lattice $\Gamma$ is trivial. 
The numbers $F^{ijk}_{lmn}$ are the $6j$ symbols, part of the algebraic data of a TQFT, $i^*$ denotes the irreps dual to $i$.
Changing the orientation of an edge is equivalent to changing its label $i$ to its dual $i^*$. 
Levin and Wen use a different normalisation than that of 
(\ref{tvpartf}), (\ref{tvpartfb}):
\beq 
d_n \left\{\begin{array}{lll}i&j&m\\k&l&n\end{array}\right\}=F^{ijm}_{kln}\ .
\label{norm}
\eeq
Before proceeding, let us write down an important algebraic property of the F symbols:
\beq \sum_{n=0}^N 
F^{mlq}_{kp^*n} F^{jip}_{mns^*} F^{js^*n}_{lkr^*}=F^{jip}_{q^*kr^*} F^{riq^*}_{mls^*} \label{BEid} \eeq
This identity is called the Biedernharn-Elliot identity or pentagon equation, which holds in every 
TQFT. In the concrete examples mentioned above, they can be proved by the definition of the $6j$ symbols as connecting
the two different fusion channels of recoupling irreps (graphically encoded by (\ref{fusion}) in section 4.2)\footnote{ by means of 
using two different ways of coupling five irreps.}.

%%%%%%%%%%%%%%%%%%%%%%%%%%%%%%%%%%%%%%%%%%%%%%%%%%%%%%%%%%%%%%
\subsection{Reconstructing 3d geometry}
%%%%%%%%%%%%%%%%%%%%%%%%%%%%%%%%%%%%%%%%%%%%%%%%%%%%%%%%%%%%%%

In our work \cite{prev} we recovered a three dimensional Turaev--Viro invariant \cite{db,turaev} from the algebra of Levin and Wen.
We associated geometric tetrahedra to the algebraic $6j$ symbols, where the edges are decorated with irreps from the $6j$ symbols.  
In that a convention needs to be adopted, e.g. the upper row should correspond to a (triangular) 
face of the tetrahedron and labels in the same column should 
correspond to opposite edges. In the examples we are looking at there is always a normalization of the $6j$'s such that they
have the same symmetry as the tetrahedron. Orientation of edges can also be taken care of in a consistent manner, we shall 
however omit them for most of what follows. Now we can translate the Biedernharn-Elliot identity
to geometry. The following picture arises:

\begin{center}\includegraphics[width=8cm]{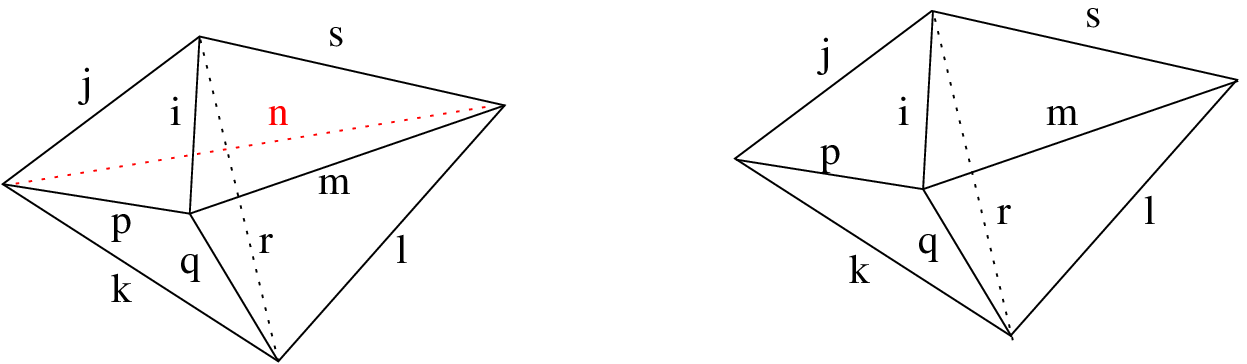}\end{center}

where the two configurations (three tetrahedra joined at the edge $n$ and two tetrahedra
glued along the triangle $(i r q)$) correspond to the left--hand and righthand side
of  (\ref{BEid}), respectively.
This is a cornerstone in proving  triangulation independence of the amplitudes 
(\ref{tvpartf})(\ref{tvpartfb})  and shows that whenever tetrahedra are glued, 
labels corresponding to internal edges have to be summed  over. 

Now, one can try to find the geometric counterpart of the operator (\ref{mag}). Constructing the dual (triangle) 
graph $\tilde{\Gamma}$ of the honeycomb lattice $\Gamma$ such that a dual edge inherits the label of the original 
edge it corresponds to (recall that edges of the original and the dual graph are in 1-1 correspondence 
in two dimensions), we proved the equality \cite{prev}:
\beq \langle \Gamma_1^{\{j\}}|\prod_p B_p|\Gamma_0^{\{j''\}}\rangle=Z_{TV}[S\times [0,1], \tilde{\Gamma}^{\{j\}}_0, 
\tilde{\Gamma}^{\{j''\}}_1]\ . \eeq
The lhs. means the matrix element of the operator $\prod_p B_p$ between two spin nets, that is, the honeycomb lattice 
$\Gamma_0\equiv\Gamma$ decorated by labels $\{j\}$ and another copy of $\Gamma_1\equiv\Gamma$ decorated by $\{j''\}$. 
The rhs. is the Turaev Viro amplitude (\ref{tvpartfb}) of the three dimensional
manifold $S\times [0,1]$ with fixed triangulations on the two boundaries given by the dual graph $\tilde{\Gamma}_i$ with the labels 
inherited from $\Gamma_i$. In figure a) below, the dashed lines show a part of $\tilde{\Gamma}_0^{\{j\}}$.
Let us concentrate on the middle vertex in the figure corresponding to the dual triangle $abc$. There is an $F$ symbol
corresponding to that vertex from all three operators $B_{p}$ of the three plaquettes sharing that vertex. To each $F$ we
associate a tetrahedron and they induce the internal triangulation of $S\times [0,1]$ 
depending on the order how the $B_p$ operators are multiplied one after the other. 
These different orders of multiplication correspond to different decompositions of the prism (built from translating the
triangle $abc$ in $\tilde{\Gamma}_0$ to the corresponding one $a''b''c''$ in $\tilde{\Gamma}_1$) into three tetrahedra.   
The fact that they commute is nicely reflected by the independence of the TV amplitude on the internal 
triangulation\footnote{Note that for proving this equality it is necessary that the coefficients of $B_p^s$ are given by 
(\ref{mco}).}. 
The figure b) shows a part of the amplitude corresponding to 
$F_{s_1\,b' c'}^{acb} F_{s_2\,c" a'}^{b'ac'} F_{s_3\,a" b"}^{c"b'a'}$.
The labels with one bar $a',b',c'$ are to be summed over when the full
amplitude is written in accordance with the fact that the underlying edges are the internal edges 
of the triangulation of $S\times [0,1]$. Summation of the $s_i$ labels comes from the sum in (\ref{mco}). 
\bc \includegraphics[width=3.5cm]{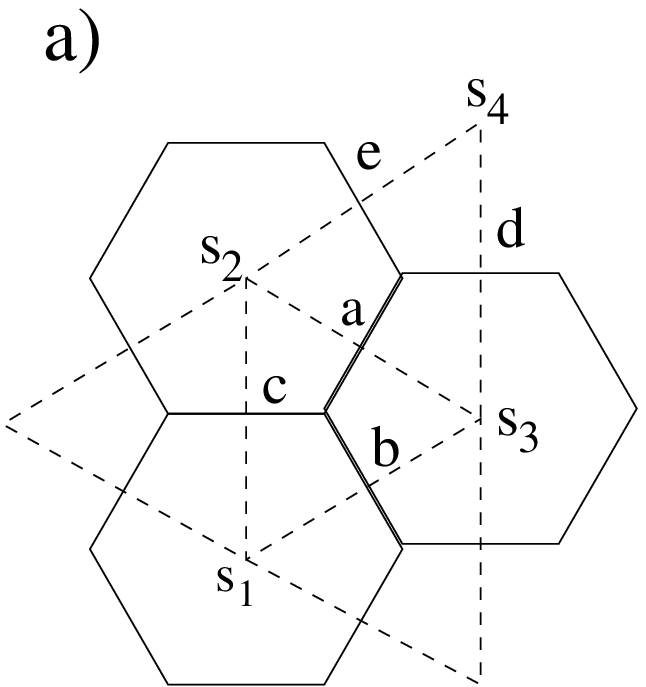}\includegraphics[width=8.5cm]{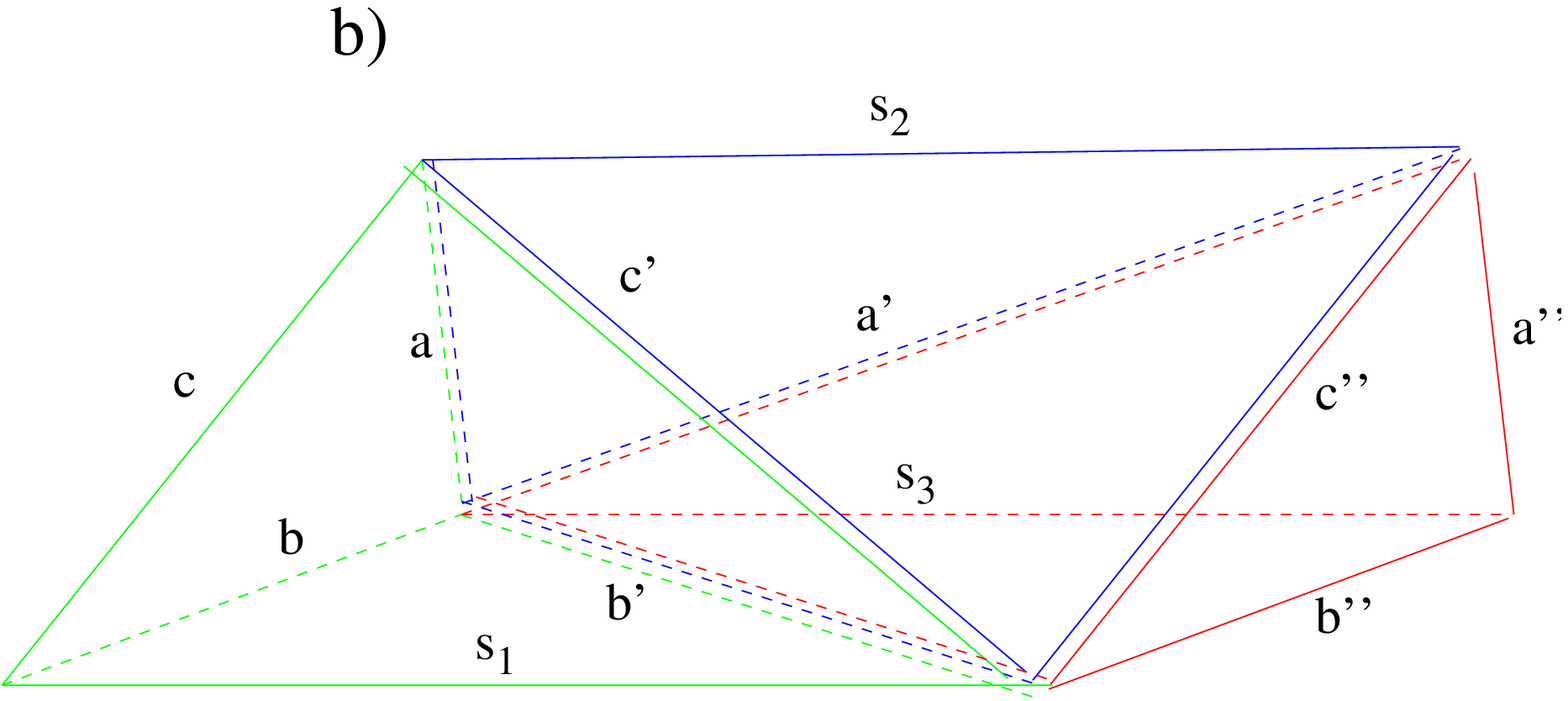} \ec
Note that $\prod_p B_p=\prod_p B_p\prod_v A_v$ if we naturally define the $6j$ symbols to be zero whenever a triple $(ijk)$ 
corresponding to a dual triangle of a tetrahedron has
$N_{ijk}=0$. This way, what we found is that the TV amplitude gives the ground state projection. For the precise matching of
all the weights to those of (\ref{tvpartfb}) and the consistency of the full amplitude, see the last section of \cite{prev}.

%Chern-Simons theories are another important class of TQFT's. Their Lagrangian is given by the Chern-Simons form  
%$CS(A)=k\,$Tr$(AdA+2/3A^3)$, where Tr is an invariant bilinear two form on the Lie algebra of the gauge group and 
%the three-form in the parenthesis is gauge invariant despite explicit appearance of the gauge field. Once the 
%normalisation of Tr is fixed, $k$ is quantized\footnote{by the single valuedness of $e^{iS}$}. Their study, 
%relation to conformal field theories is a vast field, we only want to mention one aspect. Expanding the 
%combination $CS(A+\lambda B)-CS(A-\lambda B)$ one finds a multiple of $BF+\Lambda B^3$ 

  %%%%%%%%%%%%%%%%%%%%%%%%%%%%%%%%%%%%%%%%%%%%%%%%%%%%%%%%%%%%%%%%%%%%%%%%%
  %%%%%%%%%%%%%%%%%%%%%%%%%%%%%%%%%%%%%%%%%%%%%%%%%%%%%%%%%%%%%%%%%%%%%%%%%%%%%%%%
%%%%%%%%%%%%%%%%%%%%%%%%%%%%%%%%%%%%%%%%%%%%%%%%%%%%%%%%%%%%%%%%%%%%%%%%%%%%%%%
\section{Duality and the quantum double lattice models}
%%%%%%%%%%%%%%%%%%%%%%%%%%%%%%%%%%%%%%%%%%%%%%%%%%%%%%%%%%%%%%%%%%%%%%%%%%%%%%%%

Now, we shall restrict our attention  to the case, when the TQFT is given by the structure of the double 
${\cal D}(G)$ of a finite group $G$ and show the equivalence between the lattice models of Kitaev \cite{faulttor} based on
that structure  and the corresponding string net models. Our method relies on the duality in the underlying 
lattice gauge theory \cite{kogsuss}. Essentially the same idea was employed also in the very recent paper \cite{buag}.

The Hilbert space of the Kitaev model (and that of a lattice gauge theory) is spanned by the group algebra basis 
$\{|g_1,g_2,\dots,g_E\rangle, g_i\in G\}$,
supported on an oriented lattice $\Gamma$ with $E$ edges $V$ vertices and $F$ faces. The scalar product is given by
\beq \langle g_1,g_2,\dots,g_E|h_1,h_2,\dots,h_E\rangle=\delta_{g_1,h_1}
\delta_{g_2,h_2}\dots\delta_{g_E,h_E}\eeq
%/\sum_j d_j^2
The Hamiltonian consists of two families of sums of constraint operators, which are projections and mutually commuting. We will follow
the strategy of imposing the electric constraints first and find the corresponding operators in the string net model of Levin and
Wen. Then we will study the action of the magnetic constraints in the range of the set of electric constraints, and 
determine their matrix elements in the dual basis, recovering the magnetic operators in the string net model this way.

The basic idea is the well--known expansion of any function $f\in L_2(G)$ with G being any compact Lie group 
\beq f(g)=\sum_j \sum_{m,n}^{d_j}c^{\,j}_{mn}D^j_{mn}(g)\eeq
in terms of irreps $j$ ($D^j$ are the representation matrices and $c^j_{mn}$ are coefficients). The 
statement is known as the Peter--Weyl theorem. We now define a new basis 
\beq \{|j_1,j_2,\dots,j_E,\alpha_1,\alpha_2,\dots\alpha_E,\beta_1,\beta_2,\dots,\beta_E\rangle\} \label{sn}
\eeq
by means of the scalar product
\begin{eqnarray*}&\langle g_1,g_2,\dots,g_E|j_1,j_2,\dots,j_E,\alpha_1,\alpha_2,
\dots\alpha_E,\beta_1,\beta_2,\dots,\beta_E\rangle&\\\\&={D^{j_1}}(g_1)_{\alpha_1\beta_1} 
{D^{j_2}}(g_2)_{\alpha_2\beta_2}\dots{D^{j_E}}(g_E)_{\alpha_E\beta_E}&
\end{eqnarray*}
The $\alpha_i$ ($\beta_i$) denote the target (source) index of the oriented edge $i$ and they range over the dimensions 
of the irreps $j$. We will need a linear combination of this basis defined in the following way. 
Consider all elements with fixed irreps $j_1,j_2,\dots,j_E$. For every vertex $v$ of $\Gamma$ take a 
three index--tensor $I_v$, where the indices range over the dimension of the irreps associated to the three edges (the 
honeycomb lattice is trivalent) incident to $v$. Then contract all indices with the corresponding ones in (\ref{sn}). 
A simple example corresponding to the theta--graph is given in the figure.
\bc \includegraphics[width=10cm]{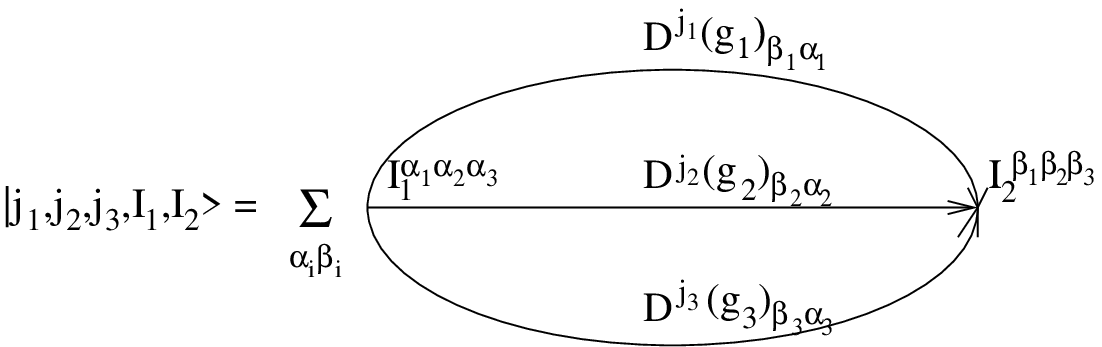} 
\ec
For these states associated to the graph $\Gamma$ we use the notation 
\beq |j_1,j_2,\dots,j_E,I_1,I_2,\dots,I_V\rangle\ .\label{spn}\eeq
%
%%%%%%%%%%%%%%%%%%%%%%%%%%%%%%%%%%%%%%%%%%%%%%%%%%%%%%%%%%%%%%%%%%%%%%%%%%%%%%%%%%%%%%%%%%%%%%%
\subsection{The electric constraints}
%%%%%%%%%%%%%%%%%%%%%%%%%%%%%%%%%%%%%%%%%%%%%%%%%%%%%%%%%%%%%%%%%%%%%%%%%%%%%%%%%%%%%%%%%%%%%%%
Let us recall the electric constraints of the Kitaev model. They are written in terms of the following operators
\[ L^g(i,v): |\dots h_i\dots\rangle\mapsto\left\{\begin{array}{ll}|\dots gh_i \dots\rangle&\mbox{if}\;i\;
\mbox{points towards}\;v,\\
|\dots h_ig^{-1} \dots\rangle&\mbox{otherwise.}\end{array}\right.\]
The local gauge transformation acting at vertex $v$ reads
\beq A_g(v)=\prod_{i\in v}L^g(i,v) \ ,\eeq
(the product is over edges incident to the vertex $v$) and the electric constraint is the projection defined as the average of the 
latter over the group: 
\beq A(v)=\frac{1}{|G|}\sum_{g\in G}A_g(v)\ \label{ec}.\eeq
Note that the range of the set of electric constraints are gauge invariant states, that is, they are invariant under 
$\prod_v A_{g_v}(v)$ with arbitrary tuple $(g_1,g_2,\dots,g_V)\in G^V$, as shown by the following calculation:
\[ A_g(v)\frac{1}{|G|}\sum_{h\in G}A_h(v)=\frac{1}{|G|}\sum_{h\in G}A_{gh}(v)=\frac{1}{|G|}\sum_{h'\in G}A_{h'}(v) \]
Hence, at each vertex, the projection (\ref{ec}) implements gauge invariance. Let us see how this is done in the general set of states
(\ref{spn}) also called spin networks. The action of a gauge transformation $A_v(g)$ on a spin network can be determined by 
rewriting the scalar product as
\beq \langle g_1,g_2,\dots,g_E|\sum_{S'}\tilde{A}^g(v)_{S,S'}|S'\rangle=\langle S|A^g(v)^\dagger|g_1,g_2,\dots,g_E\rangle\ .\eeq
Let us write down this action explicitly for a vertex $v$ whose incident edges are oriented outwards and use a simpler notation 
$|g_1,g_2,g_3\rangle$ for a generic state supported on $\Gamma$ with $1,2,3$ being the labels of the edges incident to $v$. Let us
also use a similar abbreviation $|j_1,j_2,j_3,I_v\rangle$ for $|S\rangle $ as the remaining parts are not important for the case 
at hand. Since $A^g(v)^\dagger=\prod_{i\in v}L^g(i,v)^\dagger=\prod_{i\in v}L^g(i^*,v)$ with $i^*$ denoting the opposite 
orientation for the edge $i$ we can write
\[ \begin{array}{l}{\displaystyle \sum_{\alpha_1,\alpha_2,\alpha_3} I_v^{\alpha_1\alpha_2\alpha_3}D^{j_1}(g_1)_{\alpha_1\beta_1}
D^{j_2}(g_2)_{\alpha_2\beta_2}
D^{j_3}(g_3)_{\alpha_3\beta_3} \cdots\mapsto\langle j_1,j_2,j_3,I_v|A^g(v)^\dagger|g_1,g_2,g_3\rangle}\\\\
{\displaystyle =\langle j_1,j_2,j_3,I_v|gg_1,gg_2,gg_3\rangle=\sum_{\alpha_1,\alpha_2,\alpha_3} I_v^{\alpha_1\alpha_2\alpha_3}
D^{j_1}(gg_1)_{\alpha_1\beta_1}D^{j_2}(gg_2)_{\alpha_2\beta_2}D^{j_3}(gg_3)_{\alpha_3\beta_3} \cdots}\\\\
{\displaystyle =\sum_{\gamma_1,\gamma_2,\gamma_3}\left(\sum_{\alpha_1,\alpha_2,\alpha_3}
I_v^{\alpha_1\alpha_2\alpha_3}D^{j_1}(g)_{\alpha_1\gamma_1}
D^{j_2}(g)_{\alpha_2\gamma_2}D^{j_3}(g)_{\alpha_3\gamma_3} \right) \cdot}
\end{array} \label{DDD} \]
$$
D^{j_1}(g_1)_{\gamma_1\beta_1}D^{j_2}(g_2)_{\gamma_2\beta_2}
D^{j_3}(g_3)_{\gamma_3\beta_3}\cdots =
\sum_{\gamma_1,\gamma_2,\gamma_3} I_v^{g\,\gamma_1\gamma_2\gamma_3}
D^{j_1}(g_1)_{\gamma_1\beta_1}D^{j_2}(g_2)_{\gamma_2\beta_2}
D^{j_3}(g_3)_{\gamma_3\beta_3} \cdots
$$
In the above the dots $\cdots$ stand for the remaining part of the spin network, which is not affected. In the fourth equality the
group homomorphism property of the matrices ($D(gh)=D(g)D(h)$) were used. The last equality is the definition 
of $I^g_v$ as the quantity in the big parenthesis. 

The transformation rule of $I_v\to I_v^g$ means that $I_v\in j_1\otimes j_2\otimes j_3$. The Clebsch-Gordan series 
$i\otimes j=\sum_k N_{ijk}\,k$ shows that (after a suitable unitary transformation) $I_v$ is block diagonal with
$j_1\otimes j_2\otimes j_3=\sum_{kl}N_{j_1j_2k}N_{kj_3l}\,l$ and each block transforms according to an irreps $l$ of $G$. 
The case $I_v^g=I_v,\,g\in G$ for all $g\in G$ corresponds to the trivial representation, which appears in the block decomposition iff 
$N_{j_1j_2j_3}\neq 0$. We see now that gauge invariance at
vertex $v$ is achieved by acting with the projection that projects into the invariant subspace of the decomposition. 
This should corresponds
to the projection of Levin and Wen. Assuming $N_{ijk}<2$ for every triple of irreps ensures that there is 
one unique gauge invariant tensor $I$ (for coupling three irreps), an intertwiner, so that the invariant subspace of 
$\prod_v A_v^{g_v}$ is spanned by 
\[ \{|j_1,j_2,\dots,j_E\rangle: N_{j_i,j_k,j_l}=1\;\;\text{for all}\;\;(i,k,l)\;\;\text{incident to a vertex}\}\] 
and $I$ is understood to be at every vertex contracting all $\alpha,\beta$ indices of (\ref{sn}).
For these state associated to the honeycomb graph $\Gamma$ and only for these, we will use the notation $|S\rangle$.

A shorter way to arrive at invariant spin network states is to consider a generic gauge invariant state supported 
on $\Gamma$ in the group algebra basis. These are the so--called cylindrical functions $\Psi\in L^2(G^E)$ with the invariance 
property
\beq\begin{array}{l}\langle \Psi|g_1,g_2,\dots,g_E\rangle\equiv \Psi(g_1,g_2,\dots g_E)=\\\\
\Psi(h(t_1) g_1 h(s_1)^{-1}, h(t_2) g_2 h(s_2)^{-1},
\dots,h(t_E) g_E h(s_E)^{-1})\end{array}\eeq
for every $(h_1,h_2,\dots,h_V)\in G^V$ where $t(i)$ ($s(i)$) denotes the target (source) vertex of the edge $i$. 
It can be shown that the spin network states constitute an orthonormal basis in this Hilbert space \cite{snob}.

%%%%%%%%%%%%%%%%%%%%%%%%%%%%%%%%%%%%%%%%%%%%%%%%%%%%%%%%%%%%%%%%%%%%%%%%%%%%%%%%
%%%%%%%%%%%%%%%%%%%%%%%%%%%%%%%%%%%%%%%%%%%%%%%%%%%%%%%%%%%%%%%%%%%%%%%%%%%%%%%%%
%%%%%%%%%%%%%%%%%%%%%%%%%%%%%%%%%%%%%%%%%%%%%%%%%%%%%%%%%%%%%%%%%%%%%%%%%%%%%%%
\subsection{The magnetic constraints}
%%%%%%%%%%%%%%%%%%%%%%%%%%%%%%%%%%%%%%%%%%%%%%%%%%%%%%%%%%%%%%%%%%%%%%%%%%%%%%

To recall the construction of the magnetic operators of the Kitaev model, we define auxiliary operators associated to pairs $(i,p)$ where $p$ is 
a face (plaquette) of $\Gamma$ and $i\in\p p$ is an edge on the boundary of $p$:
\beq T^g(j,p): |g_1,g_2,\dots,g_E\rangle\mapsto \delta_{g^{\pm 1}g_j}|g_1,g_2,\dots,g_E\rangle\ ,\eeq
where we have $g$ ($g^{-1}$) in the argument of the Dirac delta, when $p$ is to the right (left) of the edge $i$ 
oriented forward. The magnetic constraint is the special case $g={\bf 1}$ of the operator
\beq B_g(p)=\sum_{{h_i\in \partial p}\atop {h_1\cdots h_6=g}}\prod_{m=1}^6 T^{h_m}(j_k,p)\ .\label{ssstr}\eeq 
To adapt to the string net model we took $\Gamma$ to be the honeycomb lattice. After straightforward calculation 
one finds the action of $B_{\bf 1}(p)$ to be given by
\beq |g_1,g_2,\dots,g_E\rangle\mapsto \delta_{g_{p(1)}g_{p(2)}\dots g_{p(6)},{\bf 1}}|g_1,g_2,\dots,g_E\rangle\ ,
\label{magk}\eeq
whenever all edges $p(1),p(2),\dots,p(6)$ bounding the hexagonal face consecutively point to the 
counterclockwise direction. Should a boundary edge $p(l)$ point to the opposite direction, $g_{\,p(l)}$ needs to be replaced by 
$g^{-1}_{\,p(l)}$ in the
above expression. To proceed we write down the Plancherel decomposition of the Dirac delta function, which reads
\beq\delta_{g_1 g_2\dots g_6,{\bf 1}}=\sum_j d_j \mbox{tr}(D^j(g_1 g_2\dots g_6))=
\sum_j d_j \mbox{tr}(D^j(g_1)D^j(g_2)\dots D^j(g_6))\ .\eeq
Each term in the above sum is the scalar product of a spin network based on a two-valent graph, the hexagon with 
$|g_1,g_2,\dots,g_6\rangle$. This is a spin network of only bivalent vertices\footnote{The only intertwiner for 
bivalent vertex is the trivial Dirac delta connecting identical representations.} 
"evaluated" on the same group elements that appear
in the bounding plaquette $p$ of the spin network $|S\rangle$.
So we may write the action of $B_p:|S\rangle\mapsto \sum_j d_j |S,p,j\rangle$, where
the state $|S,p,j\rangle$ is a generalized spin network with double lines inside 
the plaquette $p$. The notion, used also  in \cite{levinwen}, nonetheless, still requires proper definition. Were the 
use of the local rules
\begin{eqnarray}
\Phi
\bpm \includegraphics[height=0.3in]{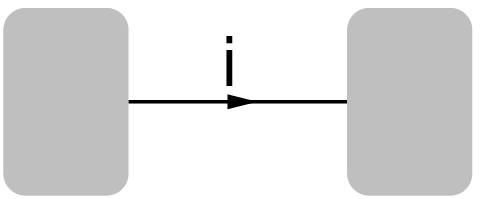} \epm  =&
\Phi 
\bpm \includegraphics[height=0.3in]{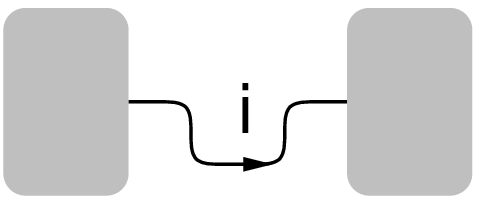} \epm
\label{topinv}
\\
\Phi
\bpm \includegraphics[height=0.3in]{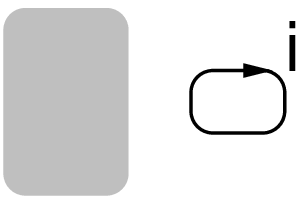} \epm  =&
d_i\Phi 
\bpm \includegraphics[height=0.3in]{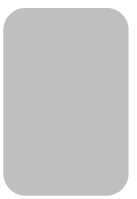} \epm
\label{clsdst}
\\
\Phi
\bpm \includegraphics[height=0.3in]{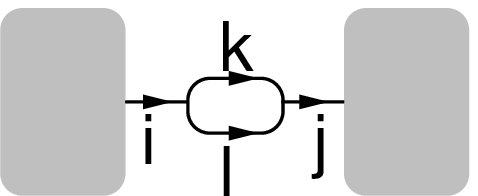} \epm  =&
\delta_{ij}
\Phi 
\bpm \includegraphics[height=0.3in]{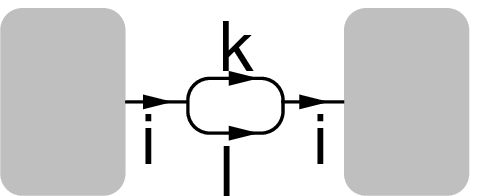} \epm
\label{bubble}\\
\Phi
\bpm \includegraphics[height=0.3in]{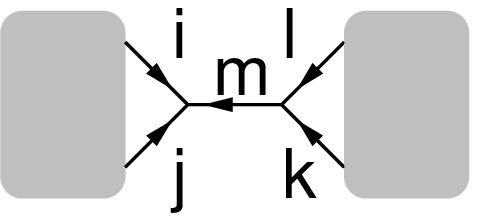} \epm  =&
\sum_{n} 
F^{ijm}_{kln}
\Phi 
\bpm \includegraphics[height=0.28in]{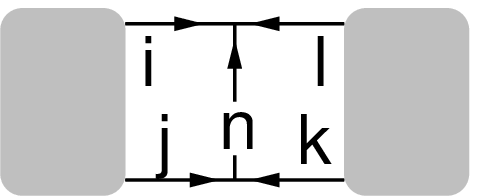} \epm
\label{fusion} 
\end{eqnarray}
of \cite{levinwen} allowed, 
we could just refer to the calculation given by formula (C1) in that article, which gives the expansion
of $|S,p,j\rangle$ in terms of bona fide spin networks $\langle S|g_1,g_2,\dots,g_E\rangle$. We could then just take it 
as the definition and we would be done. 
However, to argue in favour of these local rules in the Kitaev model, we need 
to get back to the theory in the continuum. It has been 
mentioned that in the case when the group $G$ is a Lie group, 
the electric constraints are the lattice versions of the Gauss constraint that imposes local gauge 
invariance. This was explicitly justified in the previous section. Turning to the flatness constraint: any flat connection
has trivial holonomy $g_\gamma [A]\equiv {\cal P}\exp(\int_\gamma A)$ along a 
closed curve $\gamma$ that is contractible (otherwise we could contract the 
curve to the point, whose curvature would be proportional to the generator 
of the holonomy).
The converse is also true, to every
decoration of $\Gamma$ with group elements satisfying the constraint (\ref{magk}) for all plaquettes, 
there exist smooth flat connection(s) in the manifold $\Gamma$ is embedded into.
Suppose that we have constructed one for the embedding surface of the honeycomb lattice. Then a spin network state 
\beq \Phi(S_{\Gamma'})\equiv\langle S_{\Gamma'}|g_1[A],g_2[A],\dots,g_{E'}[A]\label{ssn}\rangle \eeq
with any graph $\Gamma'$ makes sense and it is invariant of the homotopy class of the graph $\Gamma'$.
This justifies (\ref{topinv}). The connection is flat, so the holonomy along a contractible curve is ${\bf 1}$, 
Tr$D^j({\bf 1})=d_j$, which gives (\ref{clsdst}). There is no non-trivial intertwiner between two different irreps, 
whereas the lhs. of (\ref{bubble}) is a composition of invariant maps with $i\to j$ included, so that rule also holds.
Finally, we can smoothly contract the edge with label $m$ in (\ref{fusion}) without changing the value of (\ref{ssn}) and then we have
\beq \begin{array}{l}{\displaystyle \sum_{\alpha_m,\alpha'_m} 
I^{\alpha_i\alpha_j\alpha_m} I^{\alpha_l\alpha_k\alpha'_{m}} D^m({\bf 1})_{\alpha_m\alpha'_m}=
\sum_{\alpha_m} I^{\alpha_i\alpha_j\alpha_m} I^{\alpha_l\alpha_k\alpha_{m}}}=\\\\
{\displaystyle =\sum_{n,\,\alpha_n} F^{ijm}_{kln} I^{\alpha_i\alpha_l\alpha_n} I^{\alpha_j\alpha_k\alpha_{n}}
=\sum_{n,\,\alpha_n\,\alpha'_n} F^{ijm}_{kln} I^{\alpha_i\alpha_l\alpha_n} I^{\alpha_j\alpha_k\alpha'_{n}}
D^n({\bf 1})_{\alpha_n\alpha'_n}}\ ,
\end{array}\eeq
where the middle equality is a property of intertwiners and the rightmost formula coincides with the inner part of the rhs.
of (\ref{fusion}), when its middle edge with label $n$ is contracted. Note that we have omitted also the representation matrices for the
irreps $i,j,k,l$ as they are not affected by the above, as well as the other parts of the spin networks.  

Let us summarize what we have achieved. If we have a Lie group $G$ and impose gauge invariance on the honeycomb lattice $\Gamma$, 
the matrix elements of the magnetic operators in the Kitaev model in the spin network basis $\{|S\rangle\}$ can be
done in two step. First, one constructs a smooth connection in the manifold in which $\Gamma$ is embedded. 
Then uses the local rules for transforming the spin network (\ref{ssn}) as in formula (C1) of \cite{levinwen}. 
During this process, the group elements also change as we deform the edges, whose holonomies are these group elements, 
but in the end, we can deform all edges to their original location. This way we find a linear combination of the
spin network states $\{|S\rangle\}$ corresponding to the magnetic constraint given by the expression (\ref{mag}). 

Nevertheless for finite groups the local rules are, even if well motivated, postulates. The magnetic operator has 
been derived in a more direct way by introducing some auxiliary degrees of freedom in the very recent 
paper \cite{buag}.

%%%%%%%%%%%%%%%%%%%%%%%%%%%%%%%%%%%%%%%%%%%%%%%%%%%%%%%%%%%%%
%%%%%%%%%%%%%%%%%%%%%%%%%%%%%%%%%%%%%%%%%%%%%%%%%%%%%%%%%%%%%%%%%%%%%%%%%%%%%
%%%%%%%%%%%%%%%%%%%%%%%%%%%%%%%%%%%%%%%%%%%%%%%%%%%%%%%%%%%%%%%%%%%%%%%%%%%%%%%
\section{Ribbon operators}
%%%%%%%%%%%%%%%%%%%%%%%%%%%%%%%%%%%%%%%%%%%%%%%%%%%%%%%%%%%%%%%%%%%%%%%%%%%%%%%%

In section 3.1 we have been studying the ground state, the constraints that it stabilizes and the projection from the Hilbert space into
the ground state as a three dimensional TV amplitude. One of the main physical interests, however, is the string--like excitations, the
ribbon operators, which correspond to quasi--particles. We are going to sketch the corresponding preliminary results 
to illustrate that the logic 
which, worked for the ground state projection, provides us with the three dimensional
interpretation of these quantities as well.

A general ribbon in the spin net model is a string running along a certain path in the honeycomb lattice. The corresponding operator 
has the following structure
\begin{equation}
\label{strop}
W_{i_1i_2...i_N}^{i_1'i_2'...i_N'}(e_1e_2...e_N) = 
\sum_{\{s_k\}} \left (\prod_{k=1}^{N} F^{s_k}_k \right ) 
Tr \left(\prod_{k=1}^{N} \Omega^{s_k}_k \right)\ ,
\end{equation}
where $k$ runs through the vertices of the string, $e_k$ is the label of the third edge adjacent to the $k$-th vertex, 
which is not part of the string. The label $s_i$ is the "type" of the string. The index structure of the 
$6j$ symbols are given by
\begin{equation}
F^s_k = \left\{
\bmm
F^{e_{k}i_{k}^*i_{k-1}}_{s^*i_{k-1}'i_{k}'^*}\,, & 
\text{if $P$ turns left at ${\bf {I_k}}$;} \\
F^{e_{k}i_{k-1}^*i_{k}}_{si_{k}'i_{k-1}'^*}\,, & 
\text{if $P$ turns right at ${\bf {I_k}}$;}
\emm
\right.
\end{equation}
where ${\bf I_k}$ is the $k$'th vertex of the string. The $\Omega$ matrices in a string operator 
have the index structure
\begin{equation}
\Omega^{s_k}_k = \left\{
\bmm
\frac{\w_{i_k}\w_{s_k}}{\w_{i_k'}} \; \Omega^{i_{k}'}_{s_{k}s_{k+1}i_k}, & 
\text{if $P$ turns right, left at ${\bf {I_k}}$, ${\bf {I_{k+1}}}$;} \\
\frac{\w_{i_k}\w_{s_k}}{\w_{i_k'}} \; \bar{\Omega}^{i_{k}'}_{s_{k}s_{k+1}i_k}, & 
\text{if $P$ turns left, right at ${\bf {I_k}}$, ${\bf {I_{k+1}}}$;} \\ 
\delta_{s_{k}s_{k+1}} \cdot \text{Id}, &
\text{otherwise}\ ,
\emm
\right. 
\end{equation}
and generically $\Omega^{i}_{jkl}$ are matrices (so they have two more indices, which are suppressed above).
We would like to proceed as in section 3 and find a TV amplitude that a string operator describes. Before asking what the $\Omega$
matrices correspond to, let us see what geometry we find by passing the description to the dual graph and gluing a tetrahedon
whenever there is an $F$ symbol.
\begin{center}\includegraphics[width=6cm]{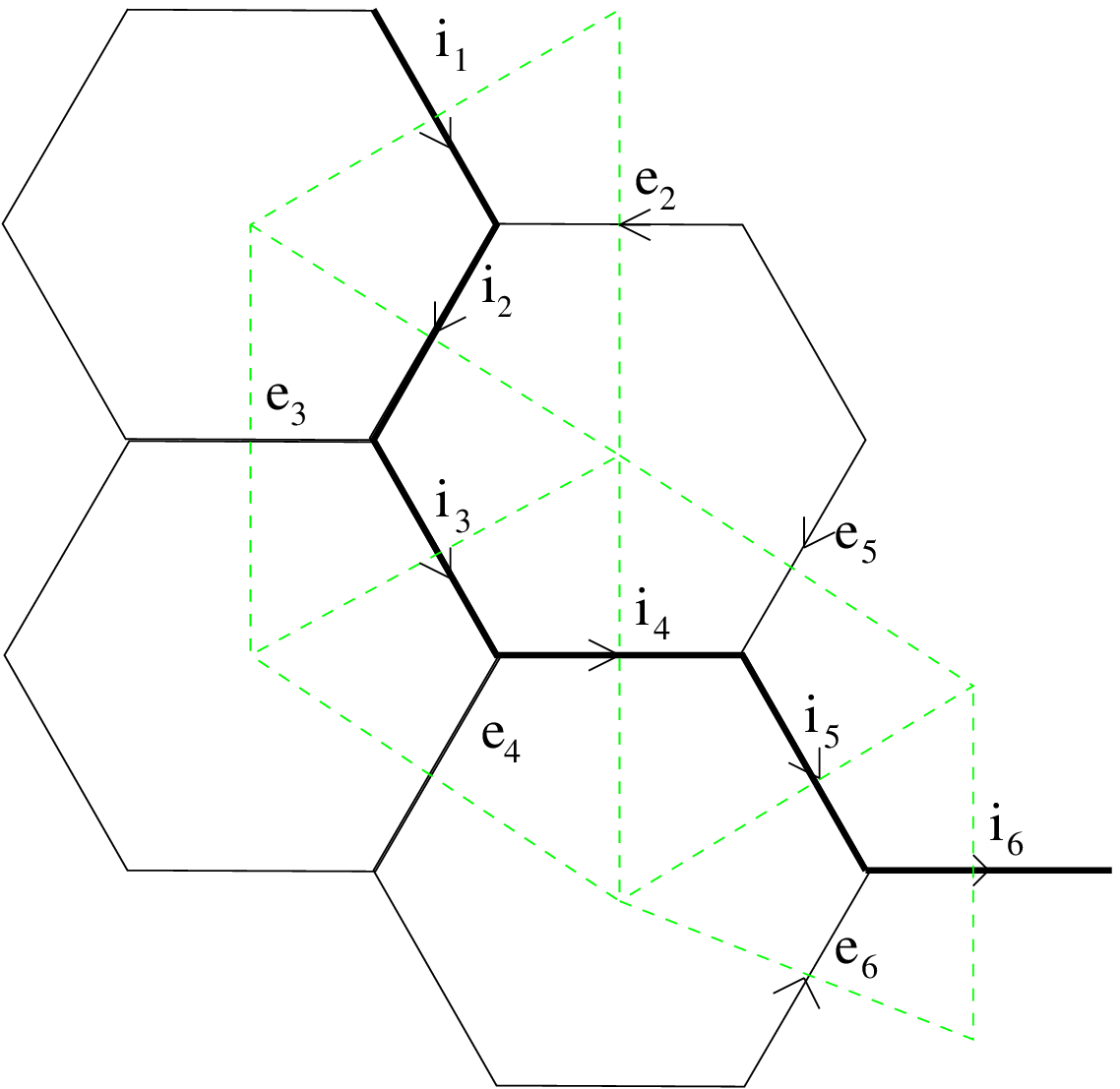}\includegraphics[width=9cm]{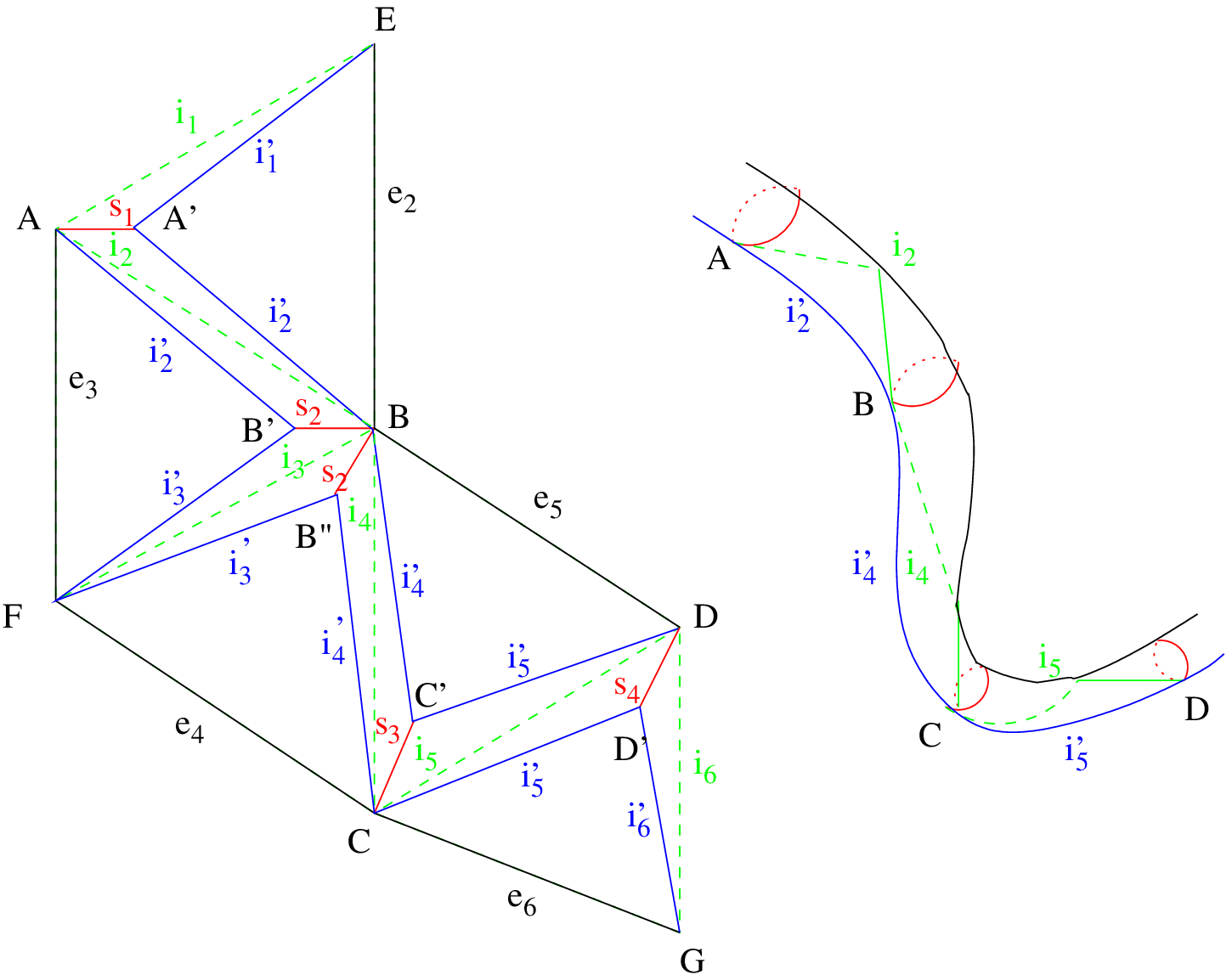}\end{center}
In the left figure we have depicted a part of a string, indicating the dual graph along. In the following we will
mean this line when referring to the string, and we will mean the collection of dual triangles 
(shown by green dashed lines in the left figure) when referring to the ribbon. In the middle 
figure we took the ribbon and drew a tetrahedron over each triangle it consists of, as dictated by (\ref{strop}).
The decoration of the edges follow the index structure of the operator. The edges with the same label belong to the same edge of the spin net, so they
are to be glued. This results in the right figure. 

We may interpret the above in the following way. There is the path $ABCD\dots$ 
in the dual graph $\tilde{\Gamma}_0^{\{j\}}$, which is
a continuous line of dual edges of the ribbon that correspond to edges of $\Gamma$, which connect vertices with
different turning directions of the string (left and right). There is an analogous path $A'B'C'D'\dots$ in 
$\tilde{\Gamma}_1^{\{j'\}}$. The gluing dictated by the algebraic structure of the operator is such that the line in 
$\tilde{\Gamma}_1^{\{j'\}}$ winds around the one in $\tilde{\Gamma}_0^{\{j\}}$ exactly once during each segment of the line.
For each such segment an $\Omega$ matrix is present in the form of the operator and  the notation
$\tilde{\Gamma}_i$ $(i=0,1$) refers to the initial and the final string net. 

The observables in the TV model are typically ribbon graphs, fat graphs or links embedded in a 
manifold, over the labels of which, there is no summation
in the amplitude \cite{db,turaev}. They are invariant under isotopy transformations. 
This property is ensured by the precise form of the braiding matrices, which then satisfy the Yang-Baxter
equation. The latter equation seems to be related to eqn. (22) of \cite{levinwen} in the spin nets. However, the precise 
relation and the identification of the braiding matrix in the TV models with an expression of $\Omega$  
as e.g. the work \cite{wittst} suggests should be found for a complete equivalence. 

%%%%%%%%%%%%%%%%%%%%%%%%%%%%%%%%%%%%%%%%%%%%%%%%%%%%%%%%%%%%%%%%%%%%%%%%%%%%%%%%%%%%%%%%%%555
%%%%%%%%%%%%%%%%%%%%%%%%%%%%%%%%%%%%%%%%%%%%%%%%%%%%%%%%%%%%%%%%%%%%%%%%%%%%%%%%%%%%%%%%%%%
\subsection{Kitaev's ribbons}
%%%%%%%%%%%%%%%%%%%%%%%%%%%%%%%%%%%%%%%%%%%%%%%%%%%%%%%%%%%%%%%%%%%%%%%%%%%%%%%%%%%%%%%%%%%
The ribbon operators are present in the Kitaev model as well \cite{faulttor}. 
\bc\includegraphics[width=8cm]{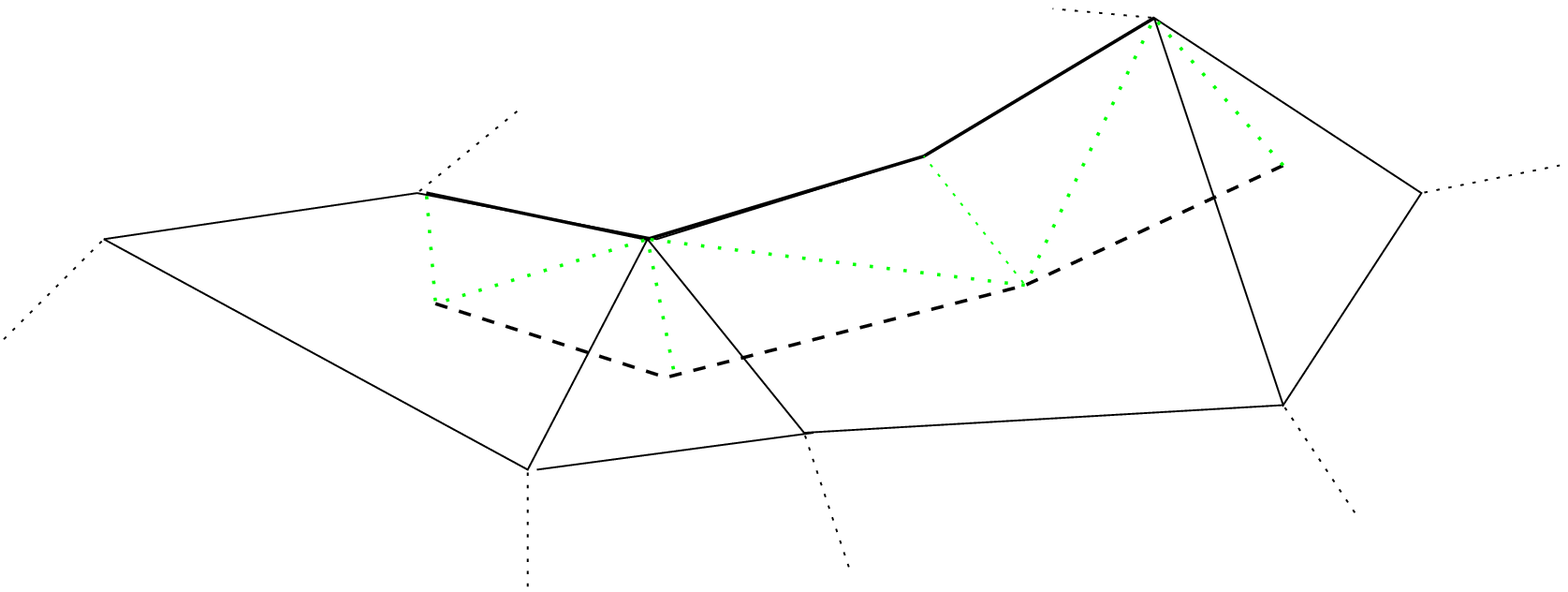}\ec
A prototypical example shown in the figure is given by a strip between a path along the edges of the original lattice 
(thick lines) and a neighbouring path in the dual lattice (dashed lines). 
It can be composed of elementary operators associated to triangles,
which connect sites, that is, pairs of a plaquette and a vertex on its boundary. In the figure sites are indicated 
by green dotted lines. One elementary building block $(i,p)$ is a triangle, which is composed of an edge $i$ and a dual 
vertex (which corresponds to a plaquette $p$). The other elementary building block $(j,v)$ is also a triangle composed of 
a dual edge (which corresponds to the original edge $j$) and a vertex $v$.   
The associated operators depend on elements of the double ${\cal D}(G)$, 
which can be represented by pairs of group elements $(g,h)$. The two types of elementary ribbon operators read
\beq W^{(h,g)}(i,v)=\delta_{g,{\bf 1}} L^h(i,v)\quad\quad W'^{(h,g)}(j,p)=T^{g^{-1}}(j,l)\ . \label{triang} \eeq
Recall that the lattice is assumed to be oriented so these formulae make sense. The composition of these elementary 
operators into a long ribbon is done by the comultiplication, which is given by
\beq W^{(h,g)}=\sum_{h_i,g_i} W^{(h_1,g_1)}W^{(h_2,g_2)}\omega^{(h,g)}_{(h_1,g_1)(h_2,g_2)}\;\;\mbox{with}\;\;
\omega^{(h,g)}_{(h_1,g_1)(h_2,g_2)}=\delta_{g,g_1g_2}\,\delta_{h_1,h}\,\delta_{h_2,g_1^{-1}hg_1}\eeq
It is desirable to express these ribbons in the spin network basis to recover their corresponding matrix elements
in the spin net model. However, there are several obstacles, which should be overcome to accomplish this task.
In \cite{levinwen}, there are additional local rules rules to reduce a generalized spin net containing ribbons, to
the basis $\{|S\rangle\}$, see the
beginning of section 4. In order for this to work, one should find a generalized spin network representation of the
above operators. Another difficulty comes about when the dual string crosses the original one. In this case, the elementary
triangles overlap and the corresponding comultiplication operations do not commute. 
One needs to find a consistent rule to define their
comultiplication in a non--ambiguous way. Note that the simplest ribbon operator in the spin net model, 
that is the one which winds around one hexagon, is easily found to correspond to (\ref{ssstr}). 
We find the following equality:
\beq B_g(p)=\sum_j tr(D^j(g))B_p^j\ .\eeq
The procedure to get it is doing the comultiplication for the six elementary operators all corresponding to the second 
type in (\ref{triang})
to arrive at $\delta_{g_1\,g_2\,g_3\,g_4\,g_5\,g_6\,g^{-1},\,{\bf 1}}$. Here the $g_i$ are the group elements corresponding to the edges in the group
algebra basis $\{|g_1,g_2,\dots,g_E\rangle\}$. Then one draws a generalized spin network representation corresponding to
the Plancherel decomposition of the Dirac delta as shown in the figure
(similarly to those for the magnetic constraints) and resolves it to the spin network basis by using the 
local moves. It is however not straightforward to generalize it\footnote{Furthermore, the argument given in section 4.2
in favour of the local rules is also lost, since the underlying connection here is not flat.}.
\bc\includegraphics[width=3cm]{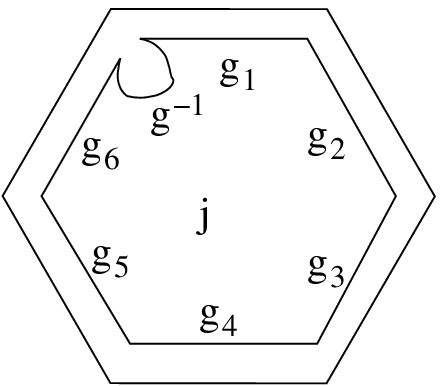}\ec

%%%%%%%%%%%%%%%%%%%%%%%%%%%%%%%%%%%%%%%%%%%%%%%%%%%%%%%%%%%%%%%%%%%%%%%%%%%
%%%%%%%%%%%%%%%%%%%%%%%%%%%%%%%%%%%%%%%%%%%%%%%%%%%%%%%%%%%%%%%%%%%%%%%
%%%%%%%%%%%%%%%%%%%%%%%%%%%%%%%%%%%%%%%%%%%%%%%%%%%%%%%%%%%%%%%%%%%%%%%%%%%%%%%
\section{Summary and outlook}
%%%%%%%%%%%%%%%%%%%%%%%%%%%%%%%%%%%%%%%%%%%%%%%%%%%%%%%%%%%%%%%%%%%%%%%%%%%%%%%%
In this paper we have been studying the lattice models of Levin, Wen and Kitaev from two perspectives. On the one hand we 
identified the ground states and the constraint operators of these models in case the underlying lattice is the honeycomb 
and the gauge group is a finite group. This has been achieved by  changing the basis from that of the group algebra,
that is, when edges are decorated by group elements, to the Fourier basis. This basis is spanned by the matrix elements of 
the irreps. A special linear combination by means of invariant intertwiners at the vertices has been shown to 
provide the range of all electric constraints and the projection at individual vertices has been identified with the
projection to the invariant subspace. Then, the magnetic operators in the group algebra basis have been shown to correspond
to those in the spin net model once the local rules postulated in the latter are satisfied. We gave an argument in favour of
them from lattice gauge theory with continuous gauge group. 

A second focus of the paper was on mapping the spin net to
the Turaev Viro state sum. We have used the idea of building up simplicial manifolds by tetrahedra with edges 
decorated with irreps corresponding to $6j$ symbols in the algebraic expressions of operators in the spin net model.      
This provided the three dimensional geometric interpretation for the ribbon operator. Having a precise TV amplitude 
identified with the ribbon operator in the spin net needs further investigation. 

One would also like to match these ribbon operators also in the model of Kitaev and the spin net of Levin and Wen. 
However, finding generalized spin network representations of the previous so that one could reduce them to the 
spin network basis is not straightforward. 

In a series of papers \cite{GaMaRa1,GaMaRa2,GaMaRa3}, families of 
$q$--deformed 'spin network automata' were implemented
for processing efficiently classes of computationally--hard problems 
in geometric topology --in particular, approximate calculations
of topological invariants of links (collections of 
knots) and of closed $3$--manifolds.
 A prominent role was played there by
`universal' unitary braiding operators associated with suitable
representations of the braid group in the tensor algebra of $(SU(2)_q\,)$. 
Traces of matrices
of these representations provide polynomial
invariants of $SU(2)_q$--colored links
(actually framed links), while 
weighted sums of the latter  give topological
invariants of $3$--manifolds presented as complements 
of  framed knots in the $3$--sphere.  These  invariants 
are  in turn recognized as partition functions and vacuum 
expectation values of physical observables (Wilson loop
operators) in $3$--dimensional 
Chern--Simons--Witten   (CSW) Topological Quantum Field Theory \cite{wittclass}.
As is well known (see e.g.  \cite{KaGoRa}, the review \cite{Oht}
and the original references therein),
any 3d TQFT of BF--type can be presented as a "double" CSW model, on the
one hand, and the square modulus of the Witten invariant
for a closed oriented 3--manifold
equals the TV invariant for the same manifold, on the other.

The remarks above make it manifest that
the efficient (approximate)
quantum algorithms proposed in \cite{GaMaRa1,GaMaRa2,GaMaRa3}
could be extended in a quite straightforward way 
to the string-net ground states and ribbon--like excitations
framed in the "naturally discretized" double $SU(2)$ CSW
environment given by the TV approach, as we have done in the present paper. 
Work is in progress in this direction.

%%%%%%%%%%%%%%%%%%%%%%%%%%%%%%%%%%%%%%%%%%%%%%%%%%%%%%%%%%%%%%%%%%%%%%%%%%%%%%%%%
\subsection*{Acknowledgements}
%%%%%%%%%%%%%%%%%%%%%%%%%%%%%%%%%%%%%%%%%%%%%%%%%%%%%%%%%%%%%%%%%%%%%%%%%%%%%%%%%%%%
Z. K. would like to thank Dirk Schlingemann and Zolt\'an Zimbor\'as for helpful discussions. 
%%%%%%%%%%%%%%%%%%%%%%%%%%%%%%%%%%%%%%%%%%%%%%%%%%%%%%%%%%%%%%%%%%%%%%%%%%%%%%%%%%%

\end{document}